\documentclass[prb,showpacs,twocolumn,amsmath]{revtex4}
\usepackage{latexsym}
\usepackage{graphicx}

\newcommand{\be}{\begin{equation}}

\newcommand{\ee}{\end{equation}}

\newcommand{\<}{\langle}

\renewcommand{\>}{\rangle}

\newcommand{\reff}[1]{(\ref{#1})}

\begin{document}

\title{Spin-Orbit Locking and Scissors Modes in rare earth crystals
with uniaxial symmetry}

\author{Keisuke Hatada$^{1,2}$, Kuniko Hayakawa$^{2,3}$, Fabrizio Palumbo$^{2}$}

\affiliation{$^1$ Instituto de Ciencia de Materiales de Arag\'on,
CSIC-Universidad de Zaragoza, 50009 Zaragoza, Spain,\\
$^2$INFN Laboratori Nazionali di Frascati, c.p. 13, I-00044 Frascati, Italy,\\
$^3$Centro Fermi, Compendio Viminale, Roma I-00184, Italy}

\date{\today}

\begin{abstract}
A recent experiment has questioned the standard relative value of spin-orbit and crystal-field strengths  in  rare-earth $4f$ electron systems, according to which  the first  should be one order of magnitude larger that the second. We find it difficult to reconcile the standard values of crystal field strength with the Single Ion Model of magnetic anisotropy. If in rare-earth systems the spin-orbit force is much larger than the crystal field, however,  spin and orbit of $4f$ electrons should be  locked to each other. For rare earths  with  non-vanishing spin,  an applied magnetic field should 
rotate both spin and charge density profile.  We suggest experiments to investigate  the possible occurrence of such  Spin-Orbit Locking, thus making a test of the standard picture, by studying the  Scissors Modes in such systems.

\end{abstract}

\maketitle

\section{Introduction}

On the basis of the standard  values of  spin-orbit and crystal-field strengths crystalline compounds fall into the following three 
categories~[\onlinecite{Skom}]:  "The relatively extended  3$d$ electrons of the transition metal-ion series  have crystal field
energies of about 1 eV as compared to spin-orbit coupling of 0.05 eV. By
contrast, rare earth $4f$-electrons are close to the nucleus, largely screened 
from the crystal field, and characterized by a dominating spin-orbit coupling of about 0.2 
eV as compared to a crystal-field interaction of the order of 0.01 eV. The magnetism caused 
by $4d, 5d$ and $5f$ electrons is intermediate, characterized by spin-orbit and 
crystal-field interactions that are both very strong". A recent paper, however,  suggests that this intermediate situation might 
be common to many, or perhaps even all, $4f$ systems~[\onlinecite{Gerr}].
We will discuss this problem in the framework of the Single Ion Model of magnetic anisotropy of systems with uniaxial symmetry. We will regard the $4f$ electron
 system
as a rigid rotor of ellipsoidal shape whose symmetry axis can precess around the symmetry axis of the cell. The action of the outer electrons is  embodied in the crystal field caused by point charge  ligands. We will find that, using the standard values of the parameters, the rotor is not polarized, due to large zero-point fluctuations which cause the average magnetism to vanish. Then assuming that the rotor  is  somehow polarized, we propose experiments to relate the spin-orbit to the crystal field strength. These experiments are based on the following property of a $4f$ electron system with nonvanishing spin if the  spin-orbit force is sufficiently strong: "The charge cloud is rigidly coupled to the spin"  so that they should rotate together under an applied magnetic field~[\onlinecite{Skom}]. We call such a structure Spin-Orbit Locking\footnotemark[1]\footnotetext[1]{The name  "rigid spin-orbit coupling" is often used in the literature on magnetism. The word "coupling", however,   might suggest to readers of different fields a property of the spin-orbit force of a particle, rather than a structure of a many-body system.}.
\begin{figure}
  \begin{center}
    \begin{tabular}{cc}
      \includegraphics[width=5cm]{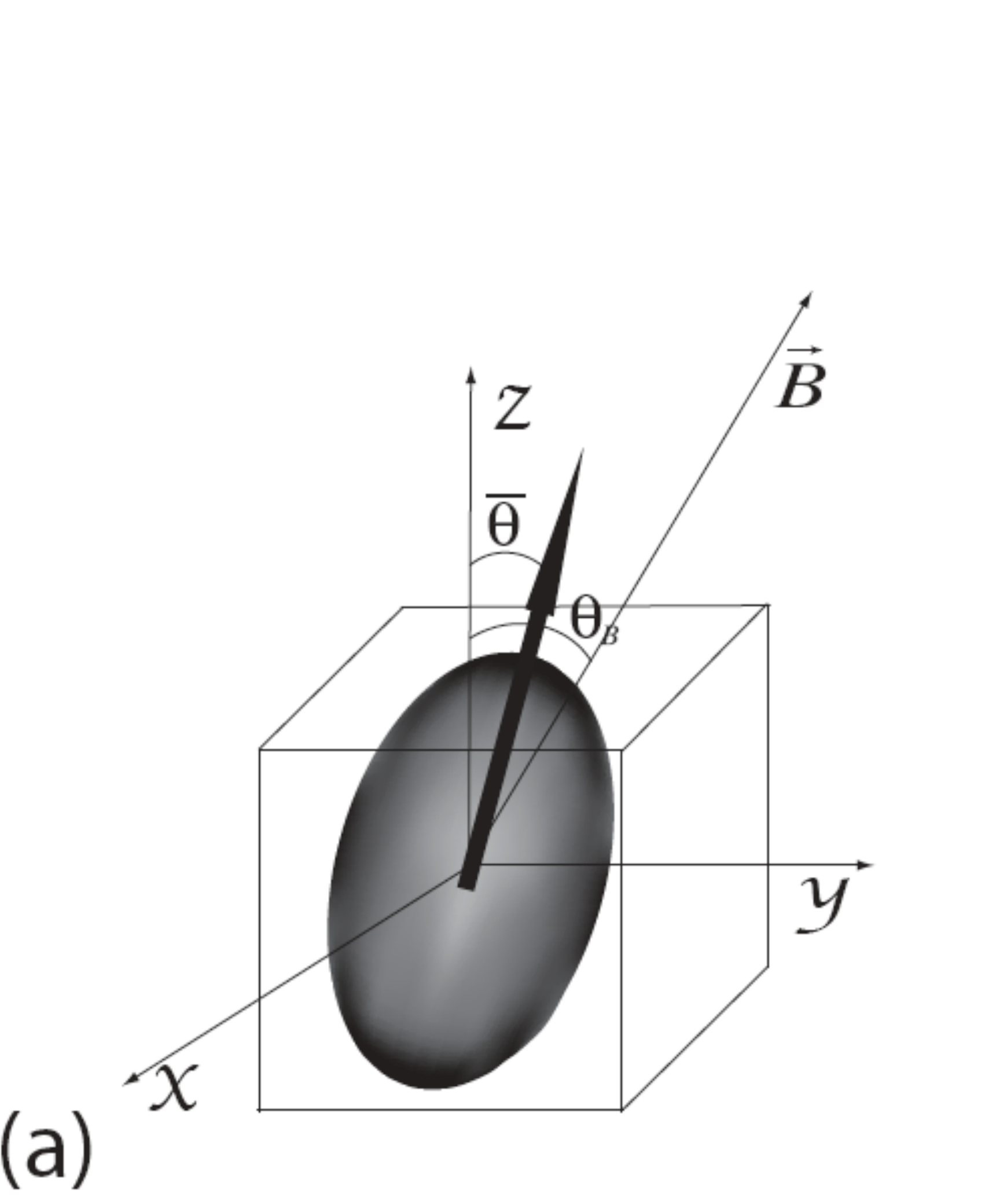}
      \includegraphics[width=5cm]{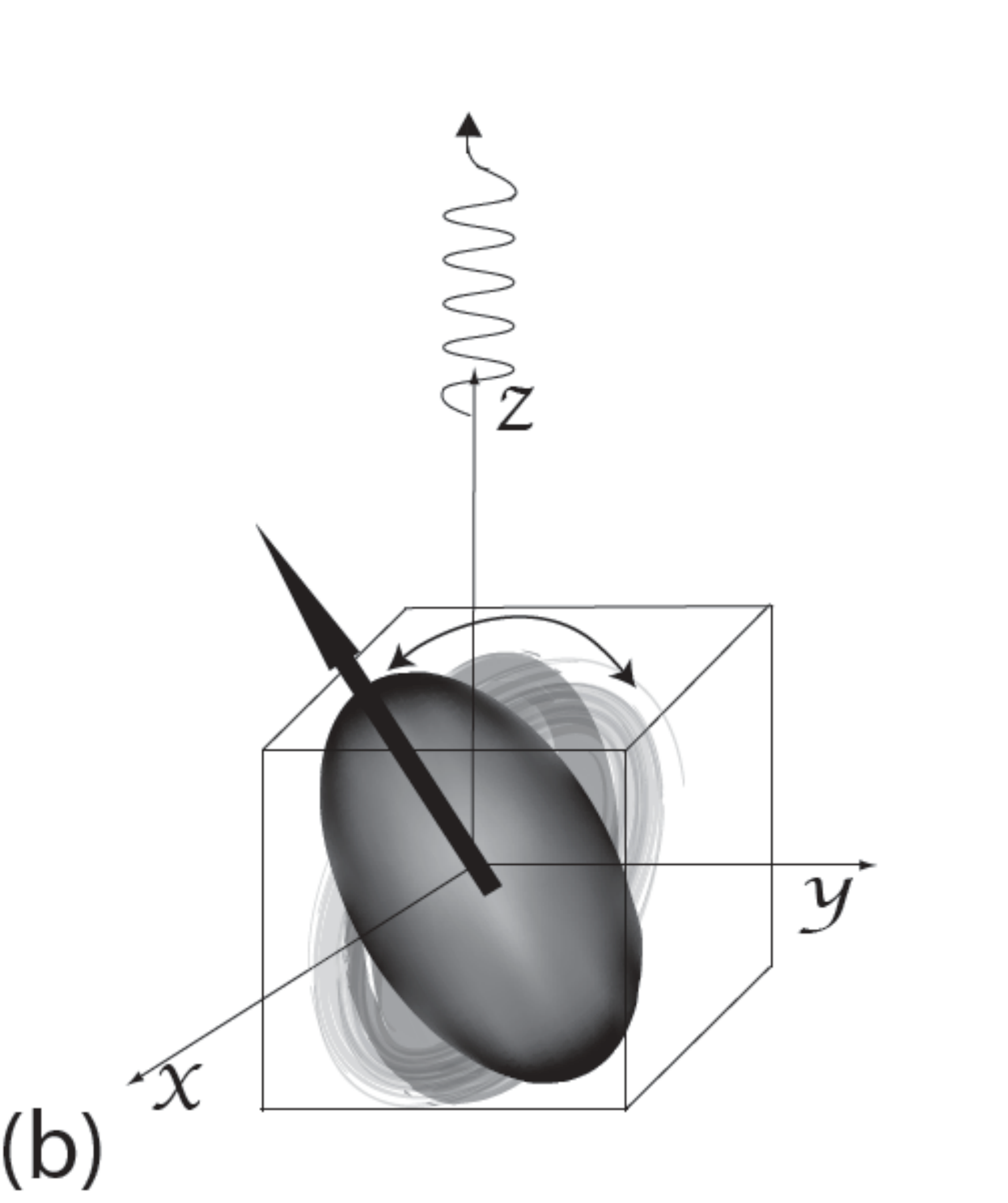}
    \end{tabular}
  \end{center}
 \caption{ The charge profile of the $4f$ electron system is rigidly coupled to the spin. In (a) the $4f$-electron system is in the ground state in the presence of the magnetic field ${\vec B}$ set at an angle $\theta_B$ with the $z$-axis. It performs zero-point oscillations  of amplitude $\theta_0$ around ${\overline \theta}$, the value of the angle at which the total potential (crystal field plus magnetic field) gets its minimum. In reality $\theta_B \ge \theta_0 >> {\overline \theta}$.
  In (b) after the magnetic field is switched off the system starts oscillating (Scissors Mode) and goes to the ground state emitting a photon.}
     \label{f1}
\end{figure}
Spin-Orbit Locking can be tested by leaving  a sample in a magnetic field which must be impulsively  switched off.  The $4f$ electron system will start oscillating and will go to the ground state emitting a photon.
The oscillating state is a  Scissors Mode, which is a kind of collective excitation  which has been predicted and observed in several many-body systems.

In Section 2 we will briefly summarize what is known about Scissors Modes and how they appear in different contexts. In particular it is interesting for us their observation in Bose-Einstein condensates by an experiment  which gave the idea of the experiments we propose, and their prediction in crystals with axial symmetry, whose physics is (assumed by us to be) essentially the same as that of the $4f$ electron system. In Section 3 we will present and discuss the Rotor Model we will use to show how Spin-Orbit Locking can be investigated studying Scissors Modes. In Section 4 we will describe the experiments designed for such purpose, in Section 5 we will discuss the life time of the state which should be created in order to observe Scissors Modes and  in Section 6 we will briefly summarize and discuss our results.

\section{Scissors Modes} 

 Scissors Modes are collective excitations in which two particle systems  move with respect to each other conserving their shape.
It was first predicted to occur in deformed atomic nuclei~[\onlinecite{LoIu}] by a semiclassical Two Rotor Model in which  protons and neutrons were assumed to form two interacting rotors to be identified with  the blades of  scissors. Their relative motion (Fig.2)
generates a magnetic dipole moment whose coupling with the electromagnetic field  provides the signature of the mode.
After its discovery~[\onlinecite{Bohle}] in a rare earth nucleus, $^{156}Gd $, and its systematic   
experimental and theoretical investigation~[\onlinecite{Ende}] in all deformed atomic nuclei, it was predicted 
to occur in several other systems including metal clusters~[\onlinecite{Lipp}], quantum dots~[\onlinecite{Serr}],
Bose-Einstein~[\onlinecite{Guer}] and Fermi~[\onlinecite{Ming}] condensates and crystals~[\onlinecite{Hata,Hata1}] (but clearly observed till now only in Bose-Einstein condensates~[\onlinecite{Mara}]). In all these systems one of the blades of the scissors must be identified
with a moving cloud of particles (electrons in metal clusters and
quantum dots,    atoms in  Bose-Einstein  and Fermi condensates, individual atoms in   crystal cells) and the other one with a structure at rest (the trap in  Bose-Einstein and Fermi condensates, the lattice in metal clusters, quantum dots and crystals). These systems can be described by a One Rotor Model.
\begin{figure}
  \begin{center}
      \includegraphics[width=5cm]{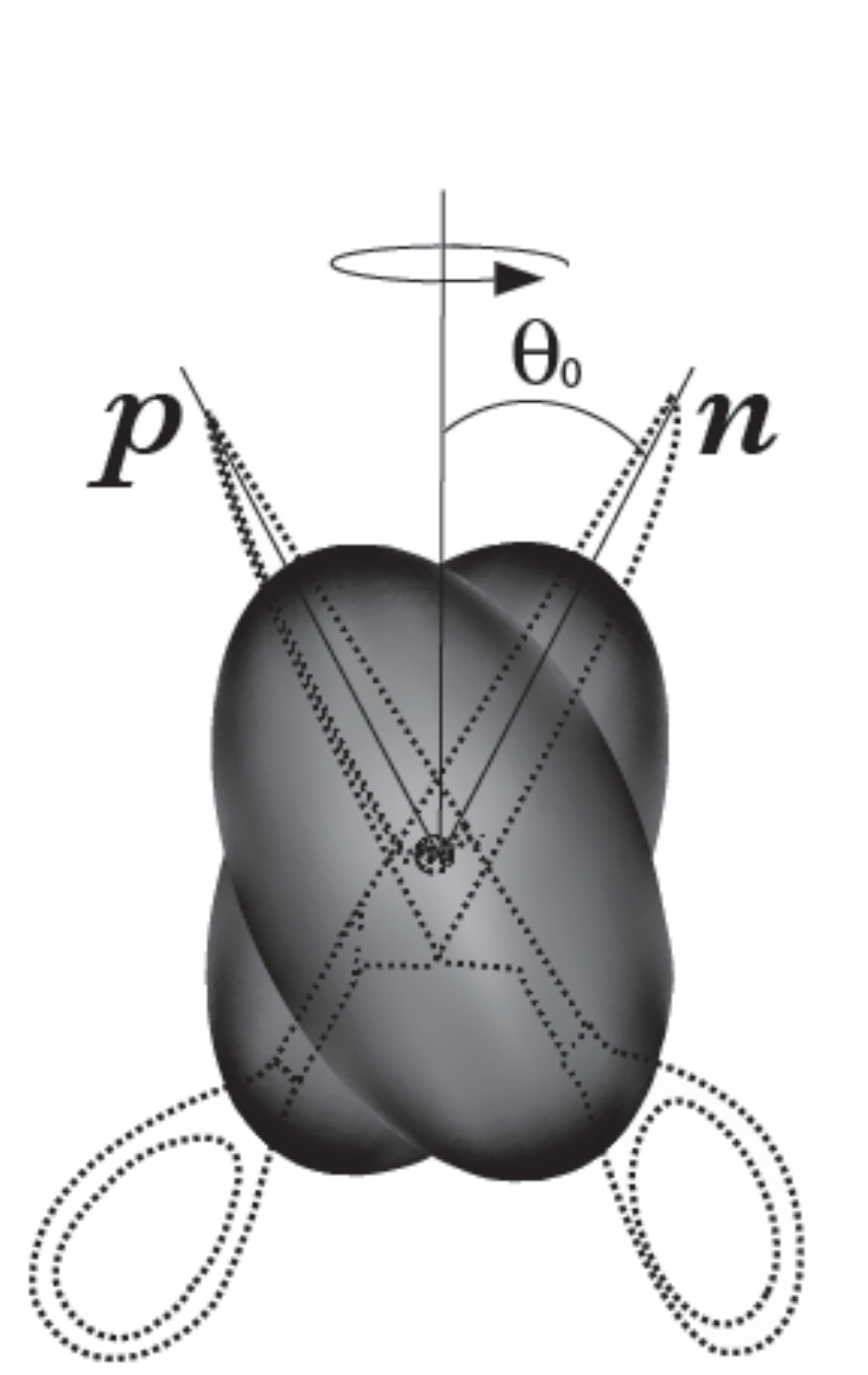}
  \end{center}
 \caption{Scissors Mode in atomic nuclei: the proton and neutron symmetry axes precess around their bisector
  }
     \label{f1}
\end{figure}
Scissors Modes in crystals have been  studied only in the framework of semiclassical models in which an atom 
 is regarded as a rigid body  which can rotate around the axes of its cell  under the electrostatic force generated by the ligands. 
  We considered crystals with  uniaxial and cubic symmetry. In the first case the precessing ion was treated as one rotor, in the second case as the body obtained by superimposing three ellipsoids at right angles. In the presence of uniaxial symmetry  {\it the photoabsorption cross section is  characterized by  a linear dichroism}~[\onlinecite{Hata}] (Fig.3). 
    \begin{figure}
  \begin{center}
   \begin{tabular}{cc}
      \includegraphics[width=5cm]{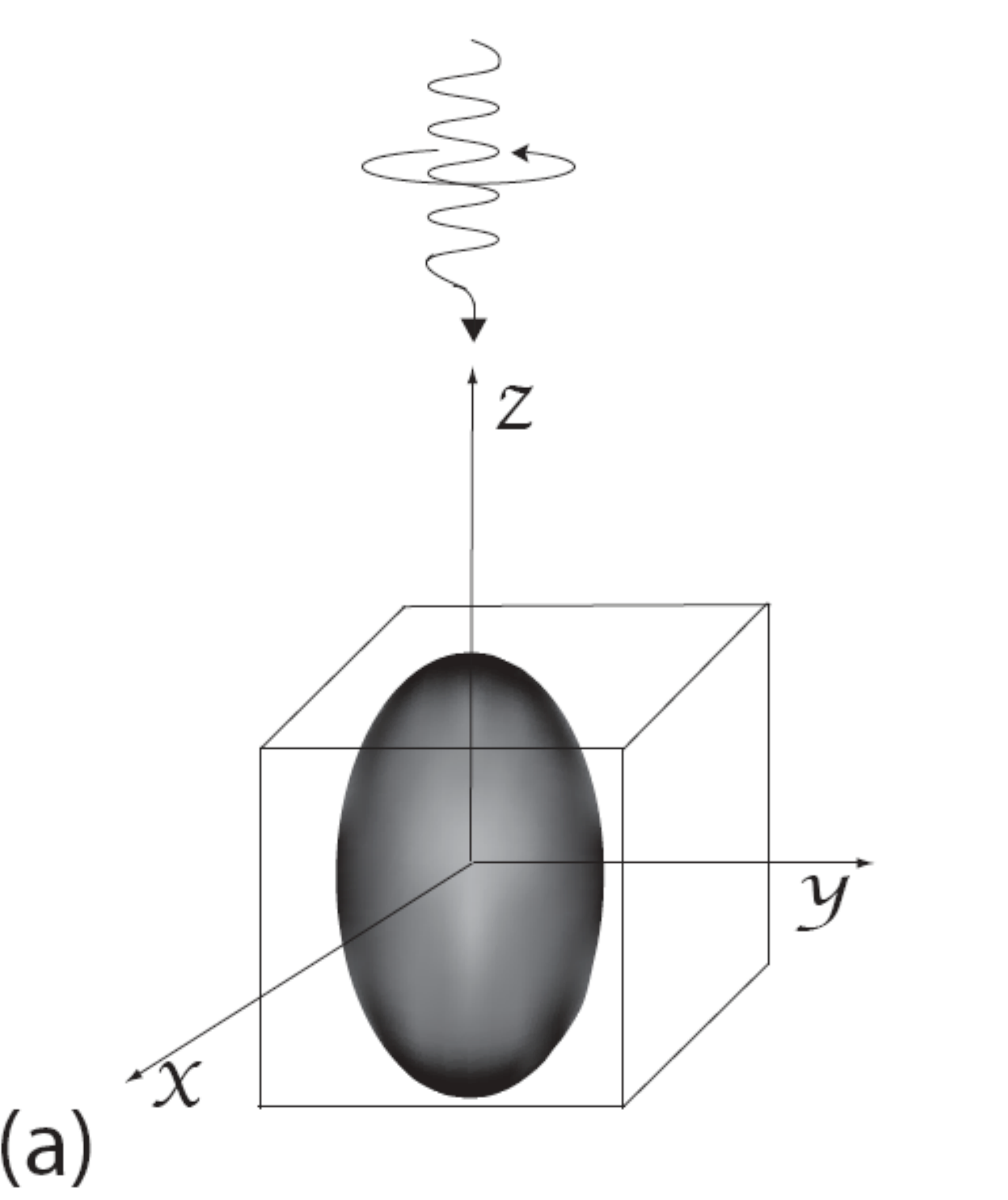}
      \includegraphics[width=5cm]{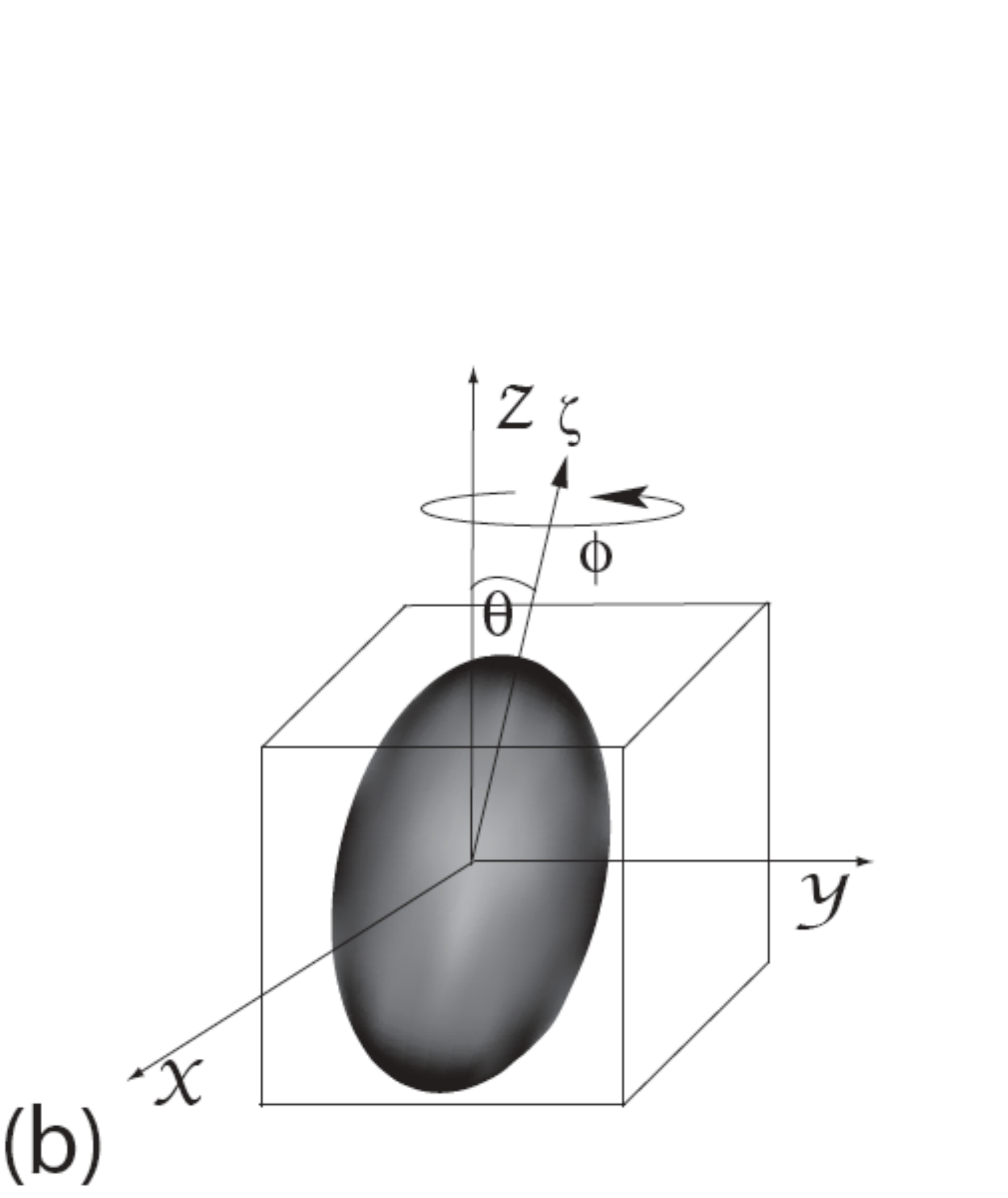}
    \end{tabular}
  \end{center}
 \caption{In a) an atom is in its ground state in a cell while a photon is incoming with circular polarization and momentum parallel to the axis of the cell. In b) the photon has been absorbed transferring its angular momentum to the atom which precesses around the symmetry axis. A photon with momentum orthogonal to the cell axis cannot be absorbed, because the atom cannot absorb its angular momentum since it cannot rotate around any axis orthogonal to the cell axis.
  }
     \label{f1}
\end{figure}
 A numerical  estimate  for LaMnO$_3$ gives a M1 transition amplitude of the order of 0.8$\AA$,  an excitation energy 
 of about 4 eV or 9 eV and a value of the zero point oscillation amplitude $\theta_0^2 \approx 0.3$ or $ 1$ depending on how the atomic moment of inertia is evaluated~[\onlinecite{Hata,Hata1}]. The uncertainty in the evaluation of the moment of inertia in a semiclassical model will be shortly discussed below.  We remark that the above values of $\theta_0$  imply a substantial polarization. With cubic symmetry  the dichroism disappears, but the values of excitation energy and M1 transition amplitude are only slightly changed.

In the present work we suggest that in a crystal in which Spin-Orbit Locking occurs Scissors Modes can be excited also by applying a  magnetic field which has an appropriate time-dependence. The mechanism of excitation would be  similar to that used with
Bose-Einstein Condensates in magnetic traps~[\onlinecite{Mara}]. In these systems one gives a sudden twist  to the trap inducing oscillations of the atomic cloud. In crystals with  Spin-orbit Locking the combined effect of the crystalline 
electrostatic field and of an external magnetic field is to create a potential well (corresponding to the magnetic trap of Bose-Einstein Condensates) which aligns  atomic spin and density profile at some angle with the direction of easy magnetization.  If we perform a sudden variation of the magnetic field (which
corresponds to twist the trap of Bose-Einstein Condensates) the atom will start oscillating around the axes of the cell and will go to the new minimum of the potential  emitting a  photon (Fig.1). As one can a priori guess
the angle by which we can rotate the atom is very small compared with the proper amplitude of Scissors oscillations, so that the amplitude of the Scissors Mode wave function in the initial state prepared by applying a magnetic field is also very small. But because of the huge  number
of cells in a macroscopic sample  an observable number of  photons of the energy of the Scissors Mode might be produced. 

If  Scissors Modes exist they   will affect the dispersive effects in the channels with their quantum numbers, which are $J^{\pi} = 1^+$. The knowledge of their  properties should then be of some importance in the study of  crystals with strong spin-orbit coupling, because  the magnetic anisotropy of these systems is at the origin of
many interesting technological applications including magnetic storage devices and sensors, spin-torque nano-oscillators for high-speed spintronics and spin-optics~[\onlinecite{Smen}]. 
In this connection we emphasize that in each of the systems studied so far, Scissors Modes  provide  specific pieces of
information. In nuclear physics they are  related to the superfluidity of deformed nuclei, in Bose-Einstein Condensates
provide a signature of superfluidity, in metal clusters they are  predicted to be responsible for paramagnetism. It is thus {\it per se} interesting, apart from being a test for Spin-Orbit Locking,  to know whether they exist also in crystals, which would add support to the idea that  they are a universal feature of  many-body systems.

\section{The rotor Model}

In  the Single-Ion Model of magnetic anisotropy[\onlinecite{Skom}] each  rare earth ion is assumed to be independent from the others. In the study of the magnetic properties only the  motion of the $4f$ electron system as a whole is considered, disregarding its excitations. In other words the $4f$ electron system is treated as a rigid rotor with spin, but, as far as we understand,  the kinetic energy of this rotor is neglected. For the application we want to do, however,  it is necessary to look at this point closely in order to understand under which conditions such an approximation can be justified.

  We   consider  the dynamics of the $4f$ electron system    with  respect to a frame of reference fixed with the cell and  $x, y, z$-axes parallel to the cell axes. We introduce  the principal frame of inertia  of this system, with axes $\xi,\eta, \zeta$. We thus introduce 6 collective  degrees of freedom,   namely the position of the  origin (which coincides with the centre of mass)  and the Euler angles $\alpha, \beta, \gamma$  of the principal frame. There remain $3  Z_{4f}  - 6 $ intrinsic position coordinates, ${\vec q}_i$ say, where $Z_{4f}$ is the number of $4f$ electrons. The explicit use of such coordinates  is terribly cumbersome for antisymmetric wave functions, but they  can be introduced  in implicit form~[\onlinecite{Palu, Sche}] if necessary.
 
 The  microscopic hamiltonian  of the $4f$ electron system can always  be written in the following way
\begin{eqnarray}
H_{4f}& = &{P^2 \over 2Zm_e} + H_{rotational}(\alpha, \beta, \gamma) +H_{intr}({\vec q}, {\vec s}) 
\nonumber\\
&+ &H_{coupl}(\alpha, \beta, \gamma,{\vec q}, {\vec s} ) +V \label{partsyst}
\end{eqnarray}
where ${\vec s}_i$ are the spins of the electrons.
The first term is the kinetic energy associated with the center of mass motion (${\vec P}$ being the total momentum and $m_e$  the electron  mass). The second term is the rotational energy of the system as a whole, the third term the energy of the electrons in their principal frame,  the forth an interaction  between rotational and intrinsic degrees of freedom, the last the crystal field potential. There is no term coupling the centre of mass coordinates with the intrinsic coordinates, because according to Galilean invariance the intrinsic motion does not depend on the centre of mass motion. On the contrary, intrinsic and rotational motion are coupled, because  the moment of inertia depends on the intrinsic motion, and because of the  centrifugal and Coriolis forces.   If the term $H_{coupl}$ is large, intrinsic excitations will disrupt the collective rotational term, and the collective Euler angles will not correspond to physical degrees of freedom. In other words, the above form of the hamiltonian is always valid  but of no practical use  in such a case. If instead $H_{coupl}$ is small and we can disregard the intrinsic excitations in the  energy range of interest,  we get the hamiltonian of a rigid rotor.   The terms $ H_{intr},  H_{coupl}$ might be studied in principle with  the methods of [\onlinecite{Palu, Sche}],
but without dwelling into such complicate analysis, we can come to a generally sound  conclusion looking at the shape of the system: if it has a well defined charge distribution which is its intrinsic  property, namely not determined by external fields, the collective approximation is generally acceptable for the lowest lying states. This seems to be the case for the $4f$ electron system of most rare earths, because they have a well pronounced quadrupole moment~[\onlinecite{Skom}]. In any case this is the approximation at the basis of the Single Ion Model, in which the intrinsic motion is altogether ignored.

For spherical rare earths, as $Gd^{3+}$ and $La^{3+}$, $H_{rotational} = H_{coupl} =0$ in Eq.(\ref{partsyst}) (the number of intrinsic variables becomes $Z_{4f}-3$). This is due to the fact that  a spherical body cannot rotate in quantum mechanics. Its spin can instead rotate, but   there is no kinetic energy associated with its  motion.  For deformed ions, instead,  the dynamics is determined by the hamiltonian
\be
H_{4f} \approx {P^2 \over 2Zm_e} + H_{rotational}(\alpha, \beta, \gamma)  +{\overline V}(\alpha, \beta, \gamma,\Sigma)  \label{collective}
\ee
where $\Sigma$ is the total spin of the system, and ${\overline V}$ results from the microscopic potential acting on the single electrons. 
{\it It is important to note  that such an approximation is generally acceptable only for the first collective excited state or at most}~[\onlinecite{Hata2}] {\it the first few ones. This observation will become of consequence in the discussion in Section 5 of the lifetime of the state prepared by applying a magnetic field to the crystal.}

It is easy to see that the fluctuations of the center of mass  are  confined within such a  small region that they can be ignored, and the center of the rotor can be assumed standing at a fixed position. The situation is in general different for the quantum fluctuations of the rotor axes. We restrict ourselves to a rotor with axial symmetry, and assume its symmetry axis along the $\zeta$-axis. Its  rotational hamiltonian is
\begin{equation}
H_{rotational}=  \frac{\hbar^2}{2{ \mathcal I}} \left( - \frac{\partial^2}{\partial \theta^2} - \cot \theta  \frac{\partial}{\partial \theta}+
\frac{1}{\sin^2\theta} L_z^2 \right) 
\end{equation}
where $\theta$ is the angle between the $z$,$\zeta$-axes,  $ L_z = - i  \hbar  \,{\partial \over \partial \phi}$ is the $z$-component of the orbital angular momentum, and ${\mathcal I}$ the moment of inertia   with respect to the $\xi$- and $\eta$-axes. For the potential we assume 
\be
{\overline V} = \frac{1}{2}C \sin^2 \theta + {\vec \mu} \cdot {\vec B}\,,
\ee
where ${\vec \mu}$ is the total magnetic moment, ${\vec B}$ the total  magnetic field acting on  the system and $C$ a restoring force constant. In the absence of magnetic field the dynamics  is determined~[\onlinecite{Paul}] by the parameter 
\be
\theta_0^2=
 { \hbar \over \sqrt{{\mathcal I} C}}\,.
 \ee
When $\theta_0 \rightarrow 0 $  {\it  the axis of the rotor can be assumed to lie along the $z$-axis, the direction of easy magnetization, and its zero-point fluctuations  can be ignored}  and.  For $\theta_0 \sim 1$ the zero-point fluctuations cannot be neglected, but the rotor is still polarized within an angle of order $\theta_0$.  {\it For  $\theta_0 >> 1$   there is no polarization at all}. 

Let us make an estimate of $\theta_0$ according to the standard values of the parameters reported in~[\onlinecite{Skom}]. 
The restoring force constant is
\be
C= 2 K_1 V_c 
\ee
 where $V_c$ is the volume of the cell and  $K_1$  the lowest order uniaxial anisotropy constant. This latter  can be expressed in terms of the second-order uniaxial crystal field parameter $A_2^0$ and of the quadrupole moment of the atom $Q_2$
\be 
  V_c K_1 = - \frac{3}{2} Q_2 A_2^0 \,.
  \ee
   For typical rare earth compounds such as  $  R_2Fe_{14}B$ and $ R_2Fe_{17}N_3$, $A_2^0 = 30 \,meV/a_0^2$ and $ -36 \,meV/a_0^2$ respectively, while for most rare earth ions  $|Q_2| \approx 0.5 \, a_0^2$ ($a_0 \approx 0.5 \, \AA$), so that $C \approx 45 \, meV$. For the present estimate of an order of magnitude we assume for the  moment of inertia  the expression appropriate to a rigid body
\be   
    {\mathcal I}_{rigid}= \frac{2}{5}m_e Z_{4f}\<r_{4f}^2\>
 \ee
     where  $\<r_{4f}^2\> $ is the mean square radius of the  $4f$-electrons.  Then setting $\<r_{4f}^2\> \approx a_0^2$ as appropriate to all $4f$-rare earth electron systems  and $Z_{4f} \approx 10$ as appropriate, for instance, to $Dy, Ho, Er$  we get
$
\theta_0^2 \approx  10
$.
By comparison we remind that the values we quoted  for $LaMnO_3$ are one order of magnitude smaller~[\onlinecite{Hata, Hata1}], and that for the atomic nuclei~[\onlinecite{LoIu}]  of the rare earths $\theta_0^2 \sim 10^{-2}$. The value of $\theta_0$ is so large because of  the small value of the restoring force constant $C$ and of the moment of inertia of the $4f$ electrons system (due to their small mean square radius).  We find it  difficul to reconcile the Single Ion Model with the standard values of the parameters, a difficulty which  might be related with the observation of Ref.~[\onlinecite{Gerr}]. Notice that increasing the crystal field strength by one order of magnitude would not alter our conclusion.

If nevertheless the magnetic anisotropy of the rare earths is due to a substantial polarization of the single ions, we might   have a direct information about  the relative  values of  spin-orbit and crystal field strength. Indeed if
the spin-orbit force remains   larger than the crystal field  Spin-Orbit Locking should occur for the rare earth ions with non-vanishing spin. In the following we suggest experiments to investigate  the   occurrence of  Spin-Orbit Locking, thus making a test of the standard picture, by studying the  Scissors Modes in such systems.

\subsection{Eigenstates and eigenvalues of the  Rotor Model}

 We set a magnetic  field in the  $y$-$z$-plane at an angle $\theta_{{\bf B}}$ with the $z-$axis,  so that its components are
 $
  { B}_x = 0, \,\,\, { B}_y= B \sin \theta_{{ B}}, \,\,\,{ B}_z =B \cos\theta_B,
$
where $B$ is its strength. 
Since the uniaxial symmetry is broken by the external magnetic field, it is convenient to introduce the cartesian coordinates
$
x=\sin \theta \cos \phi \,,   y= \sin \theta \sin \phi \,, z= \cos \theta,
$
which are the direction cosines of the axes of the atom. In the presence of a strong polarization the angle $\theta$ is very small, so that we can make the approximations
\be
x\approx  \theta \cos \phi , \,\,\,\,y \approx   \theta \sin \phi, \,\,\,\, z \approx  1-\frac{1}{2}(x^2+y^2)\,.
\ee
 The interaction with the magnetic field becomes 
 \be
 - \vec {\mu} \cdot \vec{{B}} \approx - \mu B(y \sin\theta_B + z \cos \theta_B )
 \ee
  where $\mu= g J\, \mu_B$, $\mu_B$
being the magnetic moment of the electron,  $g$  the Lande's factor and $J$ the total angular momentum of the ion.  The values of $\theta_B$ can be restricted to the interval (0, $ {\pi \over 4})$.
 Then the total potential takes its minimum at
$ x=0, 
{ \overline y} = { \mu B \sin \theta_{{\bf B}}  \over C_{{\bf B}}}
$
and the harmonic approximation to the hamiltonian   is
\begin{equation}
H_{{\bf B}} \approx  - \frac{\hbar^2}{2{ \mathcal I}} \left(  \frac{\partial^2}{\partial x^2} + \frac{\partial^2}{\partial y^2}  \right) 
+{ 1\over 2} Cx^2 + { 1\over 2} C_{{\bf B}} \, (y-{\overline y})^2 
  \end{equation}
where $C_{{\bf B}}= C + \mu B \cos \theta_{{\bf B}}$
and we neglected the constant  $- \mu \, B \cos \phi_{{\bf B}}  $. It describes the motion of the projection of the end points of the symmetry axis of the ion on the $x-y$ plane.
The eigenfunction are
\begin{eqnarray}
\psi_{n_1,n_2}^{(B)}(\phi) &=& \left( \pi \, 2^{n_1+n_2} n_1!n_2! \,\right)^{- {1\over 2}} \frac{1}{ \theta_0 }H_{n_1}\left({x\over \theta_0}\right)
\nonumber\\
&\times&H_{n_2}\left({y-{\overline y} \over \theta_0}\right)\exp \left( - {x^2 + { (y- {\overline y})^2 } \over  2 \,  \theta_0^2} \right) 
\label{eigen}
\end{eqnarray}
where $H_n$ are Hermite polynomials and 
$\theta_0^2 = {\hbar \over {\sqrt {{\mathcal I} C_B}}} \approx  {\hbar \over {\sqrt {{\mathcal I} C}}}$.

The first excited states at zero magnetic field are $\psi_{1,0}^{(0)} ,\psi_{0,1}^{(0)} $. They are analogous to the Scissors Modes of condensed atoms in magnetic traps, which can be depicted as opening and closing  scissors.  Because at zero magnetic field  the crystal field has axial symmetry, however, the eigenfunctions must be eigenfunctions of angular momentum. They  are  combinations of $\psi_{0,1}^{(0)}$ and $\psi_{1,0}^{(0)}$ 
\be
\psi_{\pm}^{(0)} = \frac{1}{{\sqrt 2}}(\psi_{1,0}^{(0)} \pm i \, \psi_{0,1}^{(0)})\,.
\ee
 All these states have  excitation energy $
 E_S \approx   \hbar  \sqrt{ {C\over {\mathcal I}}}\,.
$

It is obvious that the state $\psi_{0,0}^{(B)} $ is an excited state (but not an eigenstate) of the rotor hamiltonian at zero magnetic field. It has a life time $\tau(\psi_{0,0}^{(B)} ) $ which will be estimated in Section 5. The probability to find in  $\psi_{0,0}^{(B)} $ Scissors states is
\be
P(B) =|\langle \psi_{0,0}^{(B)} |\psi_{\pm}^{(0)} \rangle|^2 \approx  { 1 \over 2} \left({\mu B \, \theta_0^3 \, {\mathcal I} \sin \theta_B  \over \hbar^2} \right)^2 \,. \label{amplitude}
\ee
We note that the above expression has been obtained by setting the magnetic field in the $y-z$-plane at a small angle with the $z$-axis. The same result would  be obtained if it were set at a small angle with the $z$-axis in the $x,z$-plane, in which case the minimum of the potential would occur at ${\overline x} \ne 0, {\overline z}={\overline y} = 0$. On the contrary  we would get $P(B)=0$ if the magnetic field were at a small angle with the $x$- or $y$-axes, giving ${\overline x}={\overline y}=0, {\overline z} \ne 0$, because a state containing a quantum of oscillation along the $z$-axis is orthogonal to $\psi_{\pm}^{(0)}$. 

 In view of the difficulty with the rotor description  of the Single Ion Model, discussing the  experiments we will not use the restoring force constant $C$. 
 {\it We will express all the quantities  in terms of the zero point oscillation amplitude $\theta_0$,  which is unknown but must be of the order or smaller than 1 if the atom is polarized, and of the moment of inertia, for which we have a reasonable estimate}. Therefore  we will parametrize the excitation energy according to
\be
 E_S= \hbar {\sqrt \frac{C}{{\mathcal I}}} = \frac{1}{\theta_0^2} \frac{\hbar^2}{{\mathcal I}}\,. \label{energy}
\ee
 We note that
in the framework of a semiclassical model there is  no unique prescription, in general,  for the evaluation  of the moment of inertia. For instance  one might include or exclude that part of the  constituents which has a spherical shape~[\onlinecite{LoIu1,Hata1}]. In the presence of Locking such ambiguity is reduced because, qualitatively, the $4f$ electrons which contribute a spherical charge distribution have a nonvanishing spin and therefore rotate with the electrons which determine the charge deformation.

\section{The proposed experiments}

We consider 3  experiments, which are essentially 3 ways of performing one and the same experiment

  i) we set a sample of the crystal in a magnetic field which remains constant for a time sufficient for the ion in the cell to go to  the state $\psi_{00}^{(B)}$. Then   we switch  off the magnetic field  in a time not longer  than  $\tau(\psi_{0,0}^{(B)} ) $, Fig.1 
 
 ii) the magnetic field is switched on in a  time not longer than $\tau(\psi_{0,0}^{(B)} ) $  and then kept constant for a much longer time 
 
 iii) the magnetic field is pulsed with a cycle not longer than  $\tau(\psi_{0,0}^{(B)} ) $.

Let us now discuss the first experiment. After we switch off the magnetic field the state $\psi_{00}^{(B)}$ will  decay to  the ground state $\psi_{0,0}^{(0)}$ through various processes. In order  to investigate Scissors Modes we select the decay accompanied by emission of a photon of energy $E_S$. The probability of this process is $P^{(B)}$. In conclusion the  expected number of photons of energy $E_S$  is
\be
N_{photons} \, =  N_{atoms} P(B)
\ee
where $N_{atoms}$ is the  number of effective atoms in the sample, namely the atoms  whose decay photons are not absorbed in the sample itself. They are contained in a volume equal to the surface $S$ of the sample times the photon radiation length $\lambda$
\be
N_{atoms}=\rho \, \lambda \,  S
\ee
where $\rho$ is the number of atoms per unit volume. 
  We can get a lower bound to the number of radiated photons by assuming a lower bound for $\lambda$ equal to the interatomic distance which is of the order of $3 \AA$, and an upper bound  assuming $\lambda ={1 \over \sigma \rho}$, where $\sigma$ is the photoabsorption cross section. $\sigma$  has been evaluated in~[\onlinecite{Hata1}] 
\be
\sigma=6 \pi^3  {\alpha \, \hbar^2 \over m_e^2c^2}  { 1\over \theta_0^2}
\ee
where $c$ the velocity of light and $\alpha$ the fine structure constant. 
 We then get for an experiment of the first type 
\be
 3 \, S \rho \,P(B) < N_{photons} <
{ 1\over \sigma} \, S \,\rho  P(B)\,.\label{bound}
\ee
 For   a rare earth ion with $Z_{4f} =10$
 \be
 E_S \approx { 8 \over \theta_0^2} eV \,.
 \ee
If the $4f$ electrons have  $g J \approx 10$, and are in 
 a crystal of surface $S=1mm^2$ and density $\rho \approx  0.03 \AA^{-3}$, set in  a magnetic field  $B= 10 \, T$, we get 
\be
3\times 10^4 \, \theta_0^6 < N_{photons} <1.5 \times 10^8 \theta_0^8\,.
\ee

In an experiment of the second type   the initial state of the atom is $\psi_{0,0}^{(0)}$. Then we  switch on the magnetic field according to the schematic low:
\be
B(t)=
 B \,{ t\over t_1},   \,\,\, 0 < t  < t_1 \,;  \,\,\,
B(t) = B, \,\,\, t_1 \, < t \,.
\ee
Applying  standard perturbation theory to first order we get that the amplitude for the magnetic field to excite the Scissors Mode $\psi_{0,1}^{(B)}$  is
\be
- \hbar^{-1} \int_0^t dt'  e^{ i {E_S t' \over \hbar} }\<\psi_{0,1}^{(B)}| {\vec \mu} \cdot {\vec B} (t')|\psi_{0,0}^{(0)}\> \,.\label{integral}
\ee
 The  value of the matrix element~[\onlinecite{Hata1}]   is:  
\be  
\<\psi_{0,1}^{(B)}|{\vec \mu} \cdot {\vec  B} (t')|\psi_{0,0}^{(0)}\> \, \sim { 1\over 2 \theta_0} \mu \, B(t') \sin \theta_{{\bf B}} ,
\ee
and the integral in Eq.(\ref{integral}) yields
\be
\int_0^{t_1} dt  e^{ i{E_S t \over \hbar} } B(t) \approx {B \over i {E_S  \over \hbar} } \,.
\ee
In conclusion the probability per atom of exciting the Scissors Mode is
\be
\left({B \mu \over E_S}\right)^2 \left( {\sin\theta_B \over \theta_0}\right)^2 = {2 \over \theta_0^4} 
P(B)
\,. 
\ee
Since all the excited atoms will eventually decay the number of photons to be expected in the second type of experiment is
\be
N_{photons} \approx N_{atoms}  \, {2 \over \theta_0^4} \,
P(B),
\ee
which is $ 2 \, \theta_0^{-4}$ times the number of photons  in an experiment of the first type.
Finally an experiment of the third type can be analyzed in similar way. Such an experiment offers the advantage that the number of observed photons is proportional to the number of pulses of the magnetic field.

The photons emitted by the decay of the state $\psi_{00}^{(B)}$ have a signature which should help identifying them. As observed after Eq.(\ref{amplitude}) photons with  energy $E_S$ and momentum parallel to  the cell axis should be produced provided the magnetic field is set in the $x,z$-plane at a small angle with the $z$-axis,  but no photons should be produced  if if the  magnetic field were at a small angle with the $x$- or $y$-axes.

The experiments can be performed looking for photons of energy given by \reff{energy}
varying $\theta_0$ in the range $0< \theta_0 <1$. If the result is positive, we learn that both Scissors Modes and Spin-Orbit Locking exist. If the result is negative, we need to do a photoabsorption experiment to establish if Scissors Modes exist, and then Locking does not, or Scissors modes do not exist, in which case we do not learn anything about Locking.

Concerning the choice of the sample we must  exclude the ion $Eu^{3+}$ because its angular momentum  is zero~[\onlinecite{Skom}]. Moreover our analysis might apply only qualitatively to the  ions $Gd^{3+}$ and $La^{3+}$ whose 4f electron system is spherical. The hamiltonian of these ions does not have a kinetic 
term and the restoring force is of pure magnetic nature.   All  compounds with these rare earths, however, could be used by comparison, in order to exclude spurious effects.  Concrete examples of suitable compounds  can be $ R_2Fe_{14}B$ and $ R_2Fe_{17}N_3$ where the  rare earth ions can be $Dy,Ho$ and $Er$ which all have  $Z_{4f} \approx 10$ and  $gJ \approx 10$.

\section{Life time of the state $\psi_{00}^{(B)}$ }

A crucial requirement for  the  proposed experiments to be feasible is that the time required to switch on or off  the magnetic field must not be large with respect to the life time of the state $\psi_{00}^{(B)} $, $\tau\left(\psi_{00}^{(B)}\right) $, otherwise this state will go adiabatically to the state $\psi_{00}^{(0)} $ without photon emission. We need therefore an estimate of this life time. 

We notice that to order $\theta_0$ the state $\psi_{00}^{(B)} $ has a nonvanishing component  only on the Scissors Mode, so that
\be
\tau(\psi_{00}^{(B)}) =P(B)^{-1} \tau(\psi_{\pm}^{(0)}) \,.
\ee 
Using the expression of the  life time of the Scissors Modes  evaluated in Ref.[\onlinecite{Hata}] we get
\begin{eqnarray}
  \tau(\psi_{00}^{(B)}) & = &P(B)^{-1}  {4 \,  m^2 c^4\over 3 \,  \alpha} \left({ {\mathcal I} \over \hbar^2} \right)^3 \theta_0^8  \, \hbar 
  \nonumber\\
  &=&  {8\over 3}  { m^2 c^4\over  \alpha}{ {\mathcal I} \over \hbar^2} { 1\over (\mu B )^2} { 1\over \sin \theta_B} \theta_0^2  \, \hbar \,. \label{tau}
\end{eqnarray}
 For the  values of the parameters used in the estimate of the number of photons
  we get $\tau(\psi_{00}^{(B)}) \approx 3 \,\left( { \theta_0 \over \sin\theta_B}\right)^2$ sec.

There is a problem, however. As already said 
all the  excited collective eigenstates of the hamiltonian~(\ref{collective}), apart from the Scissors states,  are not physically realized, with the possible exception of a few. In order to respect  unitarity these states must be replaced by other states, which could   contribute to the width of the state $\psi_{00}^{(B)} $through single particle cascade processes. As soon as these  processes will start, they will begin to disrupt the collectivity of  
$\psi_{00}^{(B)} $ making the decay through emission of photons of energy $E_S$ impossible. 

The missing states can only be determined using the microscopic hamiltonian~(\ref{partsyst}), which is completely outside the scope of the present paper.
 In this connection, however,  we can make some considerations of general character. Before the state $\psi_{00}^{(B)} $ is substantially altered by single particle processes, a sufficient number of steps must occur. Each step  will involve one power of the relevant coupling constant. Therefore we can expect that the contribution to the width of cascade processes  will not exceed the collective contribution, provided the coupling constants are small enough and the number of steps large enough to alter significantly the structure of the state. We should feel reasonably justified in neglecting electromagnetic single particle transitions because of the smallness of the fine structure constant.   Concerning phonons,  since their energies can at most be 50 meV,  the number of steps involved should  be sufficiently large.

Anyhow the estimate (\ref{tau}) provides an upper bound to the lifetime of the state $\psi_{00}^{(B)} $. 
 
\section{Summary}

Our original motivation was to propose an experiment to excite and detect Scissors Modes in crystals, alternative to photoabsorption, in order to see if such states exist. Scissors Modes were predicted by a Two Rotor Model, which describes the motion of two interacting deformed bodies. The physics of the Two Rotor Model has been applied by many authors to  systems in which only one deformed body is moving in a nonspherical potential, which we can call a One Rotor Model.  Then we thought of using a One Rotor Model for the $4f$ electron system of the rare earths, to design an experiment by which we might investigate Scissors Modes in crystals.

In the course of this work, we read Ref.[\onlinecite{Gerr}], which questions the standard values  of spin-orbit and crystal field strength in the rare earths, and we thought that the experiments we were designing  might also provide direct  information about this issue,   namely about the existence of Spin-Orbit Locking. We stress, however,  that even if the existence of Spin-Orbit Locking (or rigid spin-orbit coupling) were regarded as firmly established, the experiments we suggest 
would have their validity: in this case they would give unambiguous information about the existence of Scissors Modes.

In the design of the experiments we adopted   the Single Ion Model of magnetic anisotropy. We then realized that, apparently,  in such a model the kinetic energy of the ion representing collectively the $4f$ electron system is disregarded, but if taken into proper account with current values of the relevant parameters for nonspherical systems, it would destroy the polarization and therefore the magnetism of the ion. This is at variance with  spherical ions, because the rotation of the spins of the latter is not associated to any kinetic energy. 

In conclusion in its present form our work contains two issues. First we raise the question of how to derive  the Single Ion Model from a microscopic hamiltonian.
Secondly, assuming that magnetism in the rare earths is however associated with single ion polarization, we propose experiments to detect Scissors Modes and investigate Spin-Orbit Locking.

\subsection *{Acknowledgment}
We are indebted to J. Chaboy and R. Natoli for invaluable discussions and to C. Masciovecchio for many  pieces of information  concerning magnetic set ups.

\end{document}